\documentclass[review]{elsarticle}

\usepackage{lineno,hyperref}
\journal{Journal of Magnetism and Magnetic Materials}

\begin{document}

\begin{frontmatter}

\title{On  magnetic structure of CuFe$_2$Ge$_2$: constrains from the $^{57}$Fe M\"ossbauer spectroscopy}

\author{Sergey L. Bud'ko}
\author{Na Hyun Jo}
\author{Savannah S. Downing}
\author{Paul C. Canfield}
\address{Ames Laboratory, US DOE and Department of Physics and Astronomy, Iowa State University, Ames, IA 50011, USA}

\begin{abstract}

$^{57}$Fe M\"ossbauer spectroscopy measurements were performed on a powdered CuFe$_2$Ge$_2$ sample that orders antiferromagnetically at $\sim 175$ K. Whereas a paramagnetic  doublet was observed above the Ne\'el temperature, a superposition of paramagnetic doublet and magnetic sextet (in approximately 0.5 : 0.5 ratio) was observed in the magnetically ordered state, suggesting a magnetic structure similar to a double-$Q$ spin density wave with half of the Fe paramagnetic and another half bearing static moment of $\sim 0.5 - 1~\mu_B$. These results call for a re-evaluation of the recent neutron scattering data and band structure calculations.

\end{abstract}


\end{frontmatter}

%\linenumbers

\section{Introduction}
The rRecent discovery of superconductivity in iron-based compounds, \cite{kam08a} followed by a flare of experimental and theoretical studies of related materials, \cite{can10a,joh10a,ste11a,wan12a} restored interest in Fe-based intermetallics with either magnetic order or enhanced Fermi-liquid properties and possible strong magnetic fluctuations, renewing, for example, interest in the RFe$_2$Ge$_2$ (R = rare earth) series. \cite{avi04a,ran11a,sub14a,sin14a,kim15a,sir15a,che16a} Among other materials the electronic structure and magnetism of CuFe$_2$Ge$_2$, \cite{zav87a} were investigated in some detail. \cite{sha15a,may16a} 

Unlike many other so-called 122 compounds, CuFe$_2$Ge$_2$ crystallizes in an orthorhombic structure (space group 51, $Pmma$), with a $2a$ site for Cu, $2d$ and $2f$ sites for Fe and $2e$ and $2f$ sites for Ge. \cite{zav87a} Band structure calculations \cite{sha15a} suggested that  CuFe$_2$Ge$_2$ has a magnetic  ground state that is ferromagnetic along $a$ direction and antiferromagnetic in other directions. Calculated magnetic moments on two Fe sites differ by less than 5\%.

Magnetization measurements in CuFe$_2$Ge$_2$ \cite{may16a}  showed an onset  of a ferromagnetic-like transition at $\approx 228$ K. On further cooling, multiple experimental techniques,   including powder neutron diffraction, \cite{may16a} identified a commensurate antiferromagnetic ordering below $T_N \approx 175$ K.   The commensurate structure was described by the propagation vector (0,~1/2,~0), so that the moments are aligned  antiferromagnetically along $b$, with chains of Fe(1) atoms ferromagnetically coupled along $a$ and antiferromagnetically coupled with Fe(2) atoms. \cite{may16a} The magnetic moments evaluated  from the  neutron diffraction data refinement were 0.36(10) $\mu_B$ on Fe(1) and 0.55(10) $\mu_B$ on Fe(2) at 135 K. An  incommensurate spin density wave  structure was reported to set in below $\approx 125$ K with a coexistence of two structures between approximately 70 and 125 K. The incommensurate structure at 4 K was described by the propagation vector ( 0,~1/2,~0.117) with magnetic moments of 1.0(1) $\mu_B$ on Fe(1) and 0.71(10) $\mu_B$ on Fe(2). The direction of the moments in both commensurate and incommensurate magnetic phases was suggested to be along the $c$-axis direction.

CuFe$_2$Ge$_2$ was identified as a metallic compound with competing magnetic ground states, that are possibly strongly coupled to the lattice and easily manipulated using temperature and applied magnetic felds. \cite{sha15a,may16a} Additionally, powder neutron diffraction data allowed for some ambiguity in the modeling of the data. \cite{may16a} All this suggested that further studies, in particular with other local probes, would be desirable to gain understanding of magnetism in this compound.

In this work we use $^{57}$Fe M\"ossbauer spectroscopy to perform a study of CuFe$_2$Ge$_2$, over a large temperature range that includes the paramagnetic and suggested magnetically ordered states.

\section{Synthesis and general characterizatoion}

Polycrystalline samples of CuFe$_2$Ge$_2$ were prepared by arc melting high purity elements on a water cooled copper hearth under $\sim 10$  mTorr of Ar atmosphere, followed by annealing.  The  weight loss after arc melting was $\sim 2$\%. The arc melted sample was put in an alumina crucible, sealed in an amorphous silica tube under a partial Ar atmosphere, and then annealed at 600$^\circ$ C for 168 hours and furnace - cooled. Given that Ref. \cite{may16a} emphasized the importance of annealing at this temperature, we made every effort to reproduce their annealing procedure.

\begin{figure}
\begin{center}
\includegraphics[angle=0,width=120mm]{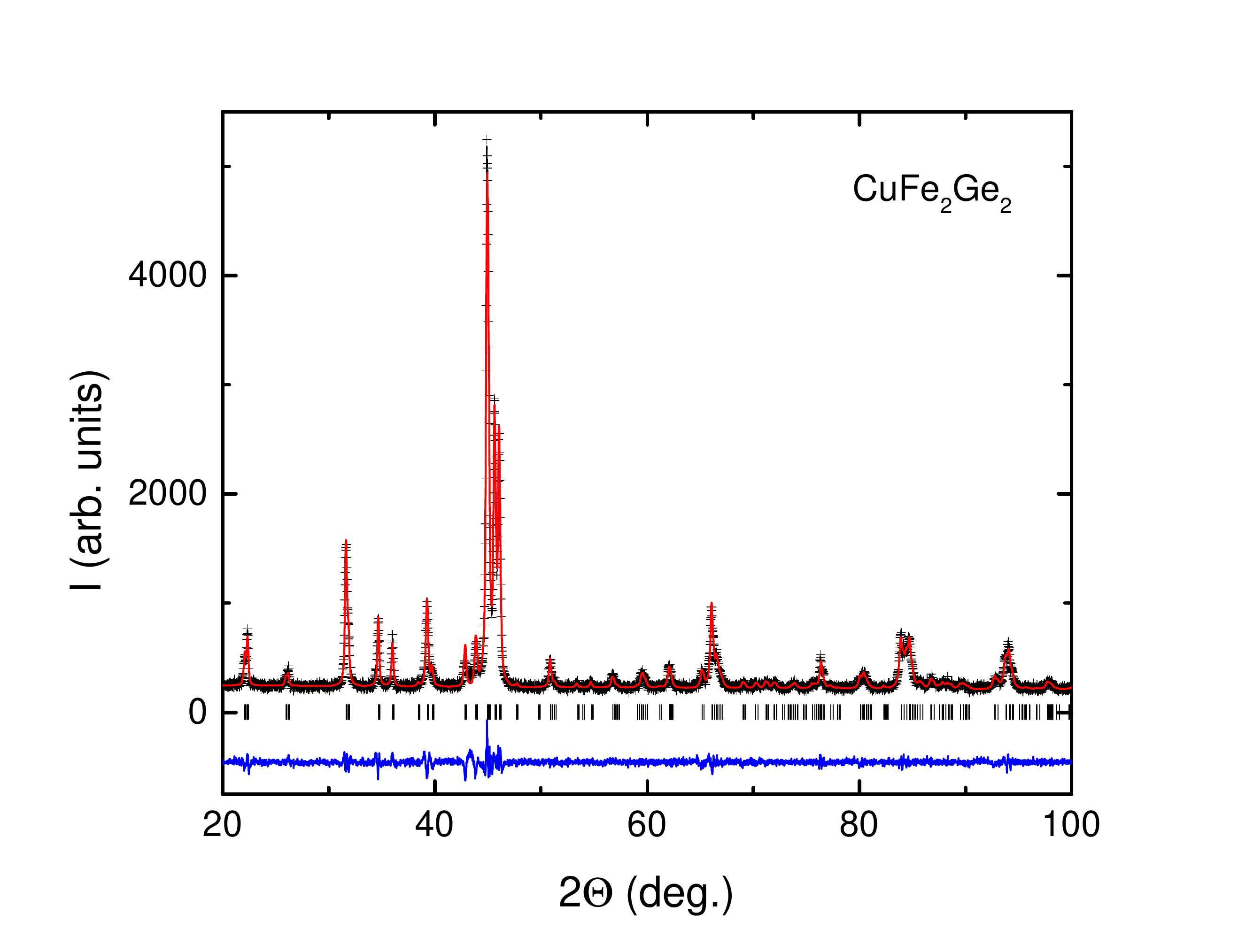}
\end{center}
\caption{(Color online) Powder x-ray diffraction data at room temperature. Data (crosses), fit (red line), calculated  peak positions (vertical bars) and the difference between experimental and calculated spectra (blue line) are shown. The refined  lattice parameters are  $a = 4.980~\AA$, $b = 3.970~\AA$, and $c = 6.795~\AA$.} \label{F1}
\end{figure}

Room temperature powder x-ray diffraction was performed using a Rigaku MiniFlex II diffractometer and zero diffraction, silicon sample holder. The results were analyzed using the GSAS software package. \cite{lar00a} The results (Fig. \ref{F1}) suggest that the sample is a single phase, the refined lattice parameters are $a = 4.980~\AA$, $b = 3.970~\AA$, and $c = 6.795~\AA$, in agreement with the literature values. \cite{zav87a}

\begin{figure}
\begin{center}
\includegraphics[angle=0,width=120mm]{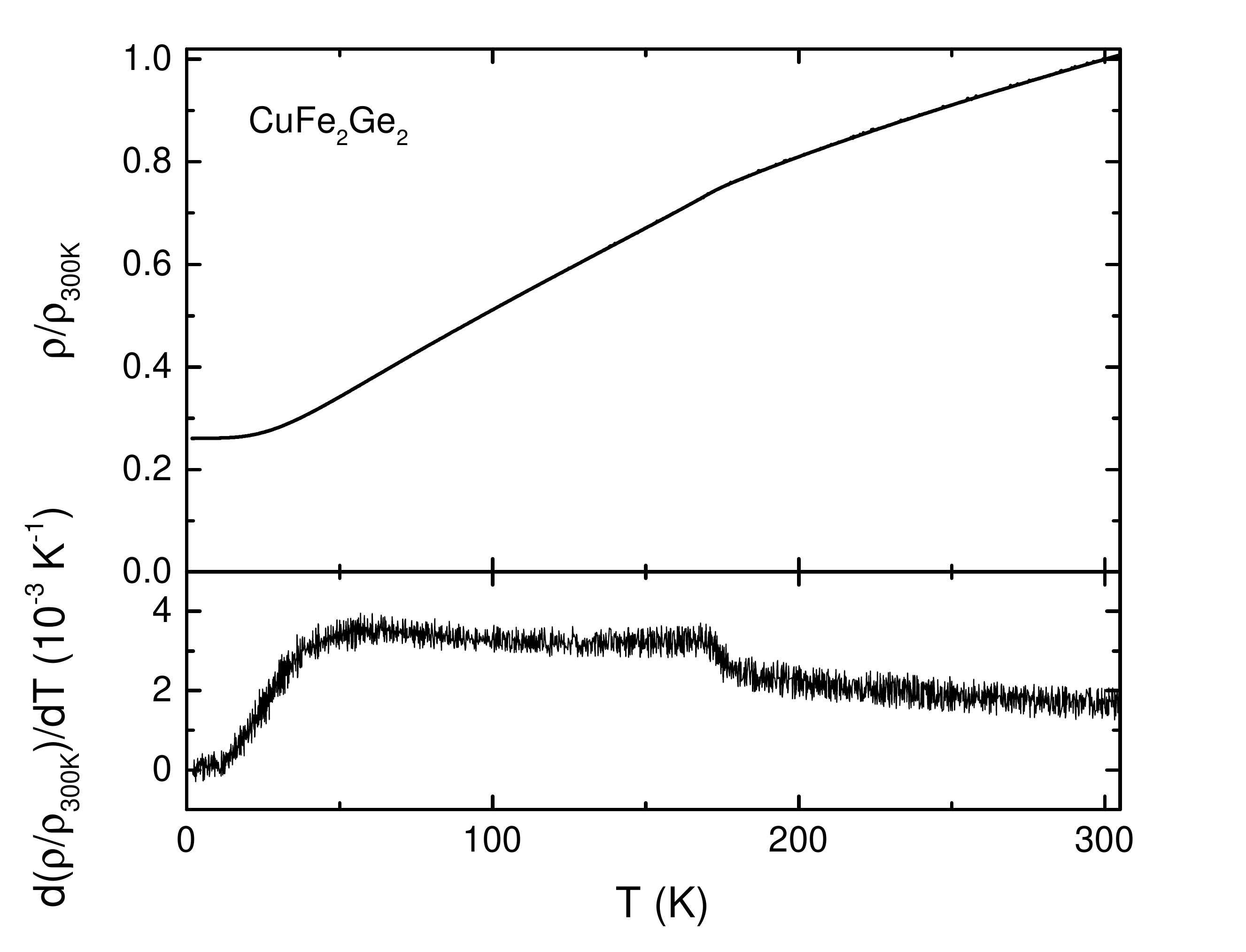}
\end{center}
\caption{Temperature-dependent resistivity of CuFe$_2$As$_2$ normalized to its value at 300 K (top panel) and its temperature derivative (bottom panel). } \label{F2}
\end{figure}

Temperature dependent resistivity measurements were performed using a conventional four-probe technique and a Quantum Design Physical Property Measurement System ACT option ($f = 16$ Hz, $I = 3$ mA). Electrical contacts to the sample were made with Epo-Tek H20E conductive epoxy and were lower than $1 \Omega$. The results of the measurements are shown in Fig. \ref{F2}. The $RRR = \rho_{300 K} / \rho_{1.8 K}$ is about 3.8. The transition at $\sim 175$ K is clearly seen both in resistivity data and in its derivative. These data are consistent with the results of Ref. \cite{may16a}, Supplementary Information.

\begin{figure}
\begin{center}
\includegraphics[angle=0,width=120mm]{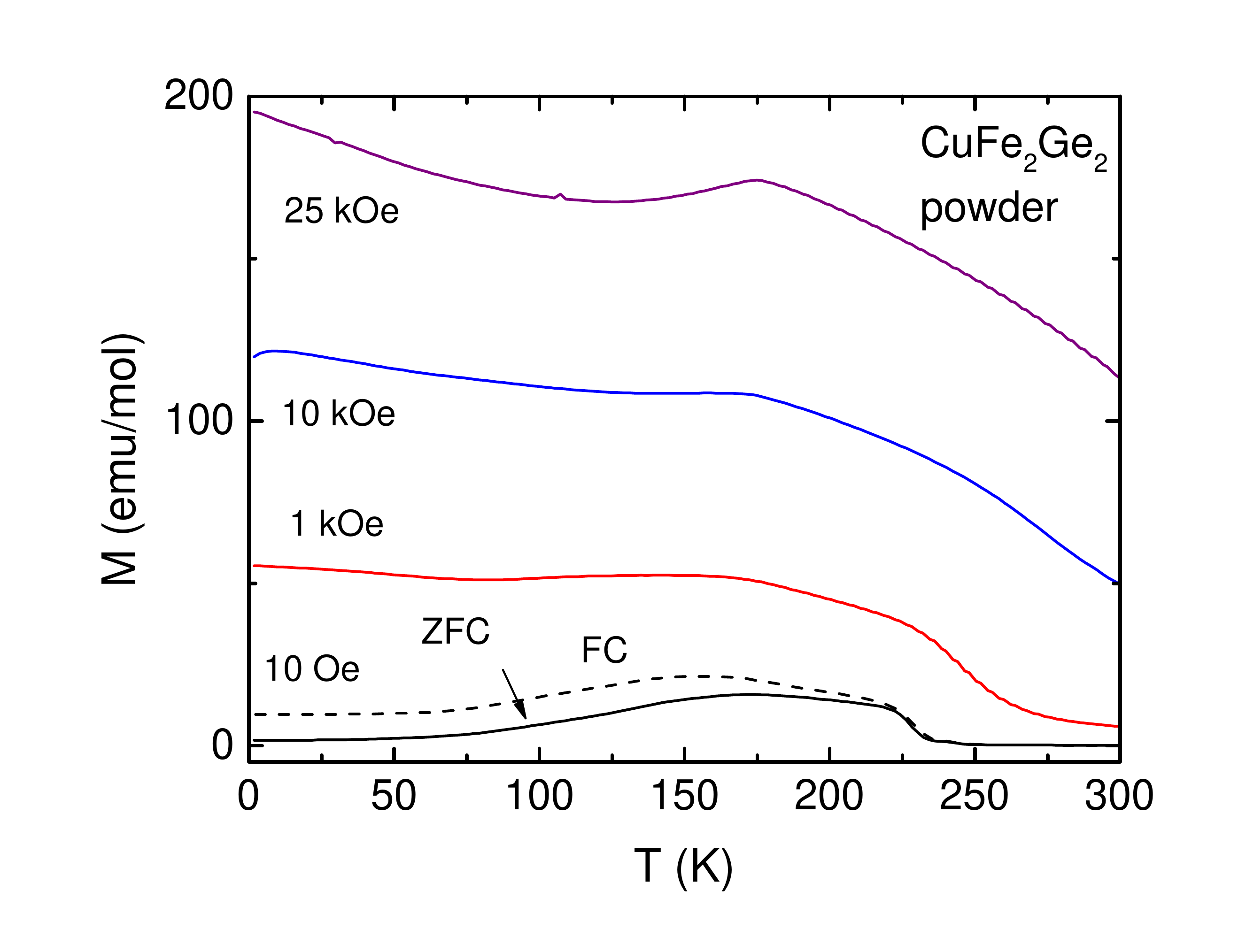}
\end{center}
\caption{(Color online) Temperature-dependent magnetization of powdered polycrystalline CuFe$_2$As$_2$ measured at four different values of applied magnetic field, 10 Oe (zero field cooled -ZFC and field cooled - FC) 1 kOe, 10 kOe and 25 kOe. } \label{F3}
\end{figure}

Temperature dependent magnetization was measured on bulk and powdered polycrystalline samples between 1.8 and 300 K for several values of applied magnetic field using a Quantum Design Magnetic Property Measurement System (MPMS 3) SQUID magnetometer. No discernible difference was observed between these two sets of data suggesting that if there is a texture (preferential orientation) in the bulk polycrystalline sample, it is insignificant. The results for powdered sample are shown in Fig. \ref{F3}. The feature associated with the transition at $\sim 175$ K is seen in all  curves. The measurement at $H = 10$ Oe suggest presence of a ferromagnetic component below $\sim 225- 230$ K, and the 1 kOe data also suggests a similar ferromagnetic component.  Distinct from the data in Ref. \cite{may16a} we do not observe any apparent feature at $T_2 \approx 125$ K and the low temperature magnetization tail in our measurements is smaller. These differences could be due either to (different) preferential orientation / texture of the polycrystalline samples, or to slightly different chemical compositions or small secondary magnetic phases possibly associated with the larger low temperature tail in Ref. \cite{may16a}.

\section{M\"ossbauer spectroscopy} 

M\"ossbauer spectroscopy measurements were performed using a SEE Co. conventional, constant acceleration type spectrometer in transmission geometry with a $^{57}$Co(Rh) source kept at room temperature. The absorber was prepared by mixing ground polycrystalline CuFe$_2$Ge$_2$ with a ZG grade (high purity) BN powder to ensure homogeneity.  The absorber was placed between two nested, white Delrin, cups of the absorber holder. The absorber holder was locked in a thermal contact with a copper block with a temperature sensor and a heater. The absorber was cooled to a desired temperature using a Janis model SHI-850-5 closed cycle refrigerator (with vibration damping). The driver velocity was calibrated using $\alpha$-Fe foil. Isomer shift (IS) values  are quoted relative to the $\alpha$-Fe foil at room temperature. The M\"ossbauer spectra were fitted using either the commercial software package MossWinn 4.0 Pre, \cite{kle16a} or the MossA package \cite{pre12a} with both analyses giving very similar results.

\begin{figure}
\begin{center}
\includegraphics[angle=0,width=120mm]{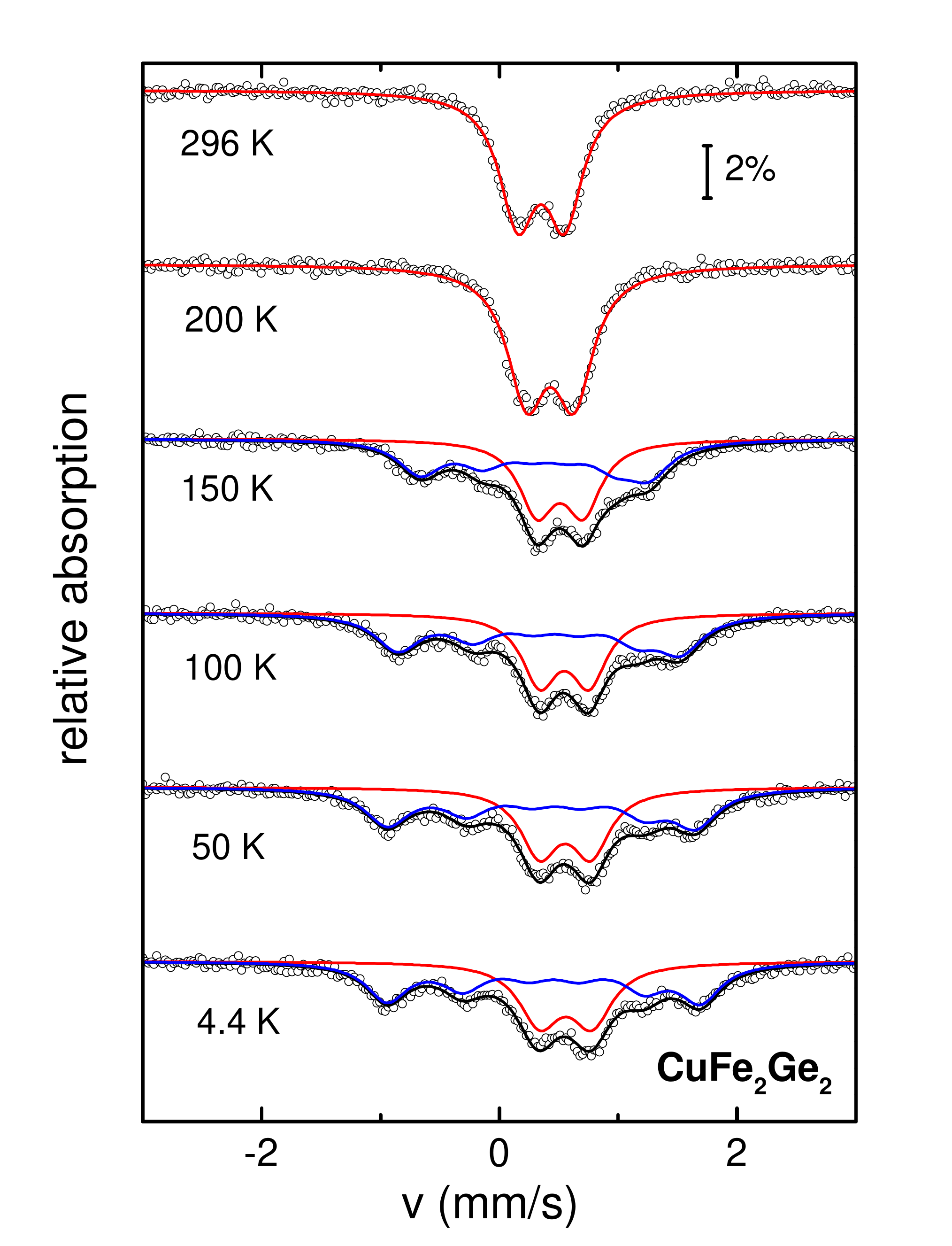}
\end{center}
\caption{(Color online) $^{57}Fe$ M\"ossbauer spectra of  CuFe$_2$As$_2$ at selected temperatures. Symbols - data, lines - fits (red - doublet, blue - sextet, black - sum of doublet and sextet). } \label{F4}
\end{figure}

$^{57}$Fe M\"ossbauer spectra at selected temperatures are shown in Fig. \ref{F4}. There is no apparent difference between the 296 K and 200 K spectra, so the feature observed in our magnetization data as well as in Ref. \cite{may16a} near $\sim 230 K$  is most probably associated with a small (below the detection level of the M\"ossbauer spectroscopy) ferromagnetic impurity. In the paramagnetic state the spectra were fitted with one doublet. Whereas this, at first glance, may appear to be at odds with two distinct Fe sites in the CuFe$_2$Ge$_2$ crystal structure,  most probably the hyperfine parameters for Fe at both sites are close enough that two separate doublets are not resolved in our measurements, instead a single doublet with  rather large linewidth is observed. 

The spectra change significantly below $\sim 175$ K (Fig. \ref{F4}). They become a superposition of a magnetic sextet and non-magnetic doublet. The results of the fits suggest that approximately half of the Fe sites bear no static magnetic moment and another half bear an ordered magnetic moment. Although it does  not follow directly from the data, is seems reasonable to  assume that Fe is magnetically ordered on one crystallographic site and not on the other. Very similar M\"ossbauer spectra were observed in so-called $C_4$ phase of e.g.  Sr$_{0.63}$Na$_{0.37}$Fe$_2$As$_2$ \cite{all16a} for which a double-$Q$ spin density wave model of the magnetic structure was suggested.

\begin{figure}
\begin{center}
\includegraphics[angle=0,width=120mm]{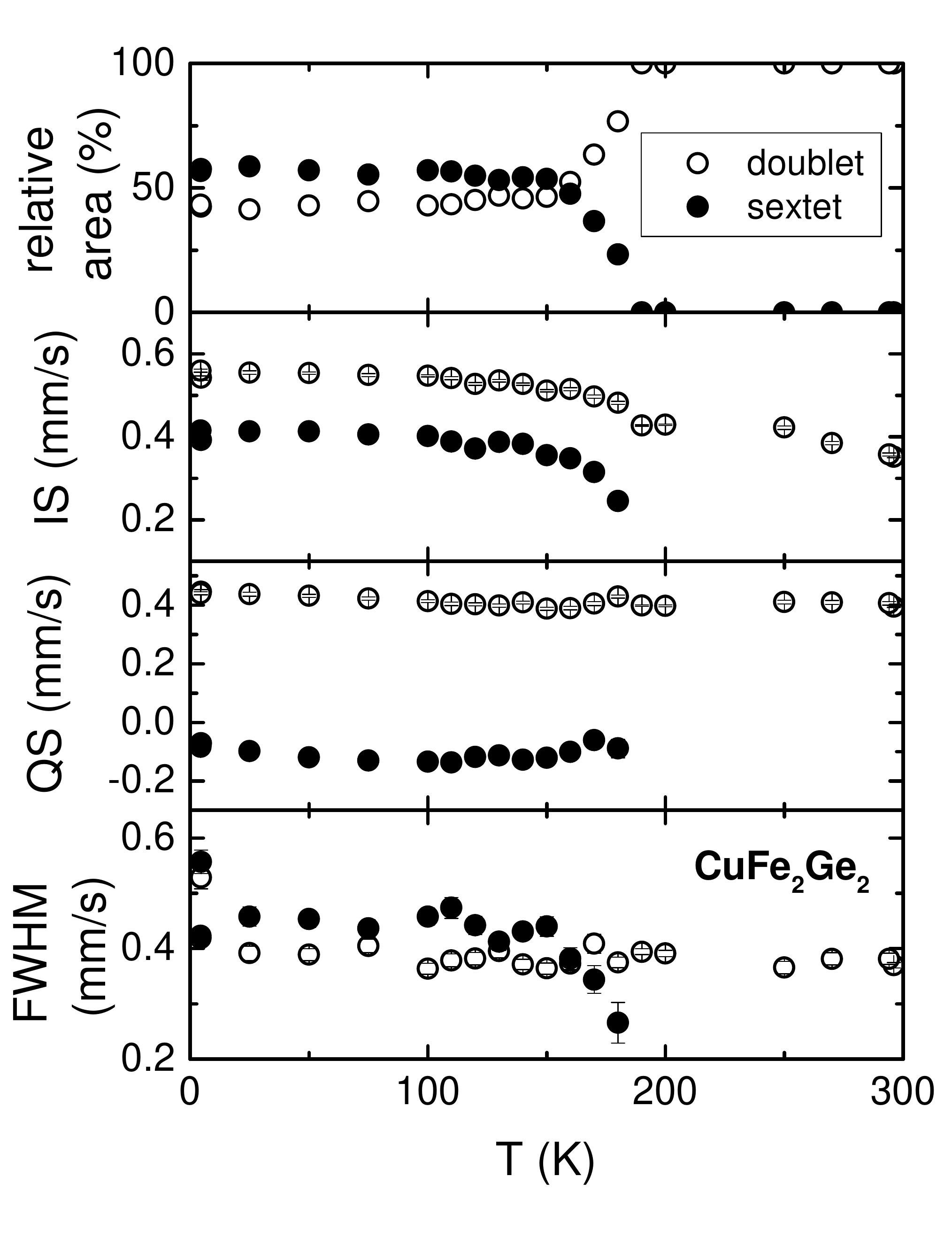}
\end{center}
\caption{Temperature dependent hyperfine parameters of CuFe$_2$Ge$_2$: relative spectral area, isomer shift (IS),  quadrupole splitting (QS), and linewidth (full width at half maximum, FWHM). Open symbols - paramagnetic doublet, filled symbols - magnetic sextet. The error bars, where not shown, are smaller than or similar to the symbols' size.} \label{F5}
\end{figure}

\begin{figure}
\begin{center}
\includegraphics[angle=0,width=120mm]{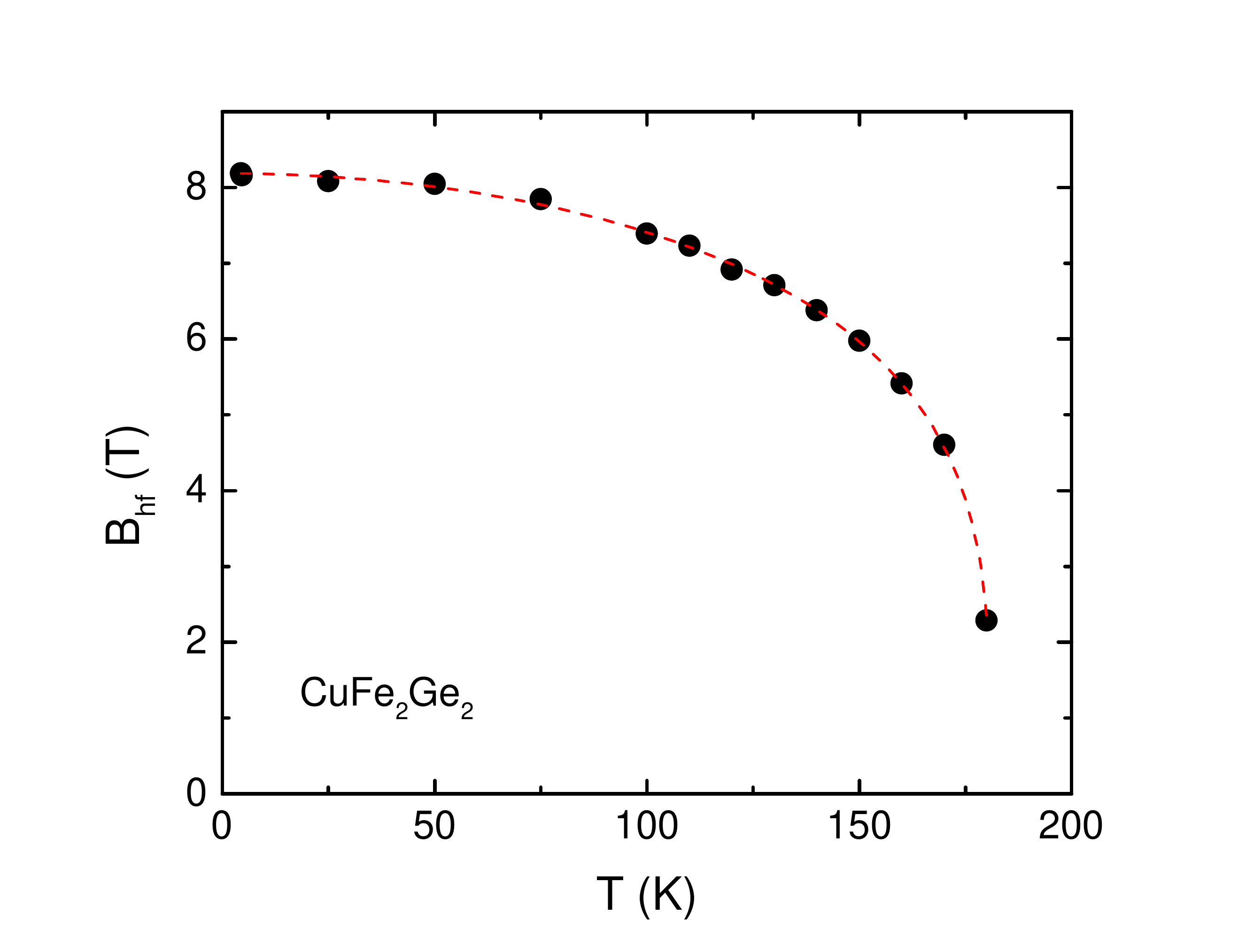}
\end{center}
\caption{(Color online) Temperature dependent hyperfine field inferred from the analysis of the sextet. Dashed line - fit with a phenomenological equation $B_{hf}(T) = B_{hf}(0) [1-(T/T_N)^{\alpha}]^{\beta}$ (see the text).  } \label{F6}
\end{figure}

The temperature dependence of the hyperfine parameters of CuFe$_2$Ge$_2$ are plotted in Figs. \ref{F5}, \ref{F6}. Both the relative area of the sextet (Fig. \ref{F5} upper panel) and the hyperfine field (Fig. \ref{F6}) are finite for $T = 180$ K. In Fig. \ref{F5}, the isomer shift (IS) values for the doublet and the sextet both increase with decrease of temperature, whereas the quadrupole splitting (QS)  has only a minor variation with temperature. The linewidth values associated with both the doublet and the sextet are rather high ($\sim 0.4$ mm/s) over the whole temperature range. In comparison with tetragonal RFe$_2$Ge$_2$ (R = rare earth) \cite{bar90a} and AFe$_2$As$_2$ (A = Ba, Sr, Ca, Eu), \cite{mcg14a} the values of the isomer shift for CuFe$_2$Ge$_2$ nonmagnetic doublet are comparable but higher, whereas those for quadrupole splitting are measurably higher, pointing to the higher anisotropy of Fe atoms environment in CuFe$_2$Ge$_2$

A closer look at the isomer shift and the quadrupole splitting associated with the doublet reveals anomalies at approximately the magnetic ordering temperature. These anomalies are not surprising since the magnetic transition is accompanied by the visible features in the $c$ - lattice parameter\cite{may16a} that can affect structural and electronic environments of the Fe atoms in the structure. 

The temperature dependence of the hyperfine field inferred from the sextet in the M\"ossbauer spectra is presented in Fig. \ref{F6}. The hyperfine field, $B_{hf}$, varies continuously and approaches zero smoothly for $T \geq 181$ K. This behavior is consistent with a second order of the magnetic phase transition. There is no discernible feature at $\sim 125$ K, where transition from commensurate to incommensurate spin density wave state was inferred from neutron scattering data. \cite{may16a} The $B_{hf}(T)$ dependence was fitted using a phenomenological form $B_{hf}(T) = B_{hf}(0) [1-(T/T_N)^{\alpha}]^{\beta}$, \cite{blu03a} where $T_N$ is N\'eel temperature, $\alpha$ and $\beta$ are parameters describing behavior for $T \rightarrow 0$ and near $T_N$ respectively. The results of the fit give $B_{hf}(0) = 8.19 \pm 0.02$ T, $T_N = 180.8 \pm 0.2$ K, $\alpha = 2.0 \pm 0.1$, $\beta = 0.27 \pm 0.01$. 

The observed hyperfine field serves as a measure of the magnetic moment on the Fe site. Although there is no unique relationship between the magnetic hyperfine field and the magnetic moment, \cite{dub09a}, for rough evaluation of the moment we can use the value of hyperfine coupling constant $A = 15$ T/$\mu_B$ for metallic Fe, that yields $\sim 0.55 \mu_B$. Since in related AFe$_2$As$_2$ (A = Ba, Sr, Ca, Eu) compounds using this value of the hyperfine coupling constant leads to underestimate of the Fe moment by approximately a factor of 2, \cite{mcg14a} more realistic evaluation for the moment on magnetic  Fe site is probably closer to $1 \mu_B$.

\section{Discussion and Summary}

The results of $^{57}$Fe M\"ossbauer spectroscopy measurements at different temperatures on CuFe$_2$Ge$_2$ confirm that at $\sim 180$ K this compound has a second order transition to a magnetically ordered phase. In contrast to the published interpretation of the neutron scattering data and to the band structure calculations that conclude that both Fe sites  in CuFe$_2$Ge$_2$ are in ordered states bearing moments of similar values, our data suggest that half of the Fe sites (possibly corresponding to a specific crystallographic site) experience no finite hyperfine field (are paramagnetic) over the whole temperature range from 4.4 K to 296 K, whereas another half have a static magnetic moment of $\sim 0.5 - 1~ \mu_B$ below the magnetic ordering temperature. The M\"ossbauer spectroscopy results call for re-eavaluation of the band structure calculations and neutron scattering data. \cite{sha15a,may16a} It is very possible that, like the AFe$_2$As$_2$, AeFe$_2$As$_2$ and AAeFe$_4$As$_4$ (A =  Ba, Sr, Ca, Eu, Ae = Cs, Rb, K) \cite{can10a,joh10a,ste11a,wan12a,iyo16a,mei17a} there is near degeneracy of multiple magnetic orderings. If this is indeed the case for CuFe$_2$Ge$_2$, then it may meet some of the necessary (albeit not sufficient) conditions for Fe - based superconductivity.

 \section*{Acknowledgements}

This work was supported by the U.S. Department of Energy, Office of Basic Energy Science, Division of Materials Sciences and Engineering. The research was performed at the Ames Laboratory. Ames Laboratory is operated for the U.S. Department of Energy by Iowa State University under Contract No. DE-AC02-07CH11358. NHJ and SSD were supported by the Gordon and Betty Moore Foundation EPiQS Initiative (Grant No. GBMF4411).

\clearpage

\clearpage

\clearpage

\clearpage

\clearpage

\clearpage

\section*{References}

\begin{thebibliography}{1}

\bibitem{kam08a}
 Y. Kamihara, T. Watanabe, M. Hirano, H. Hosono, Iron-Based Layered Superconductor La[O$_{1-x}$F$_x$]FeAs ($x$ = 0.05-0.12) with $T_c$ = 26 K, J. Am. Chem. Soc. 130 (2008) 3296-3297.

\bibitem{can10a}
P. C. Canfield and S. L.   Bud'ko, FeAs-Based Superconductivity: A Case Study of the Effects of Transition Metal Doping on BaFe$_2$As$_2$, Annu. Rev. Condens. Matter Phys.  1  (2010) 27-50.

\bibitem{joh10a}
D. C.  Johnston, The puzzle of high temperature superconductivity in layered iron pnictides and chalcogenides, Adv. Phys.   59  (2010) 803-1061.

\bibitem{ste11a}
G. R. Stewart, Superconductivity in iron compounds,  Rev. Mod. Phys.   83 (2011) 1589-1652.

\bibitem{wan12a}
N.-L.  Wang, H.  Hosono and P.-C.  Dai (eds.), Iron-based Superconductors. Materials, Properties and Mechanisms,  Pan Stanford Publishing, Boca Raton, FL , 2012.

\bibitem{avi04a}
M. Avila, S. L.  Bud'ko, and P. C.  Canfield, Anisotropic magnetization, specific heat and resistivity of RFe$_2$Ge$_2$ single crystals, J. Magn. Magn. Mater. 270 (2004) 51–76. 

\bibitem{ran11a}
Sheng Ran, Sergey L. Bud'ko, and Paul C. Canfield, Effects of substitution on low-temperature physical properties of LuFe$_2$Ge$_2$, Philos. Mag., 91 (2011) 4388-4400.

\bibitem{sub14a}
Alaska Subedi, Unconventional sign-changing superconductivity near quantum criticality in YFe$_2$Ge$_2$, Phys. Rev. B 89 (2014) 024504.

\bibitem{sin14a}
David J. Singh, Superconductivity and magnetism in YFe$_2$Ge$_2$, Phys. Rev. B 89 (2014) 024505.

\bibitem{kim15a}
H. Kim, S. Ran, E. D. Mun, H. Hodovanets, M. A. Tanatar, R. Prozorov, S. L. Bud'ko, and P. C. Canfield, Crystal growth and annealing study of fragile, non-bulk superconductivity in YFe$_2$Ge$_2$, Philos. Mag. 95 (2015) 804-818.

\bibitem{sir15a}
N. Sirica, F. Bondino, S. Nappini, I. P\'i\v{s}, L. Poudel, A. D. Christianson, D. Mandrus, D. J. Singh, and N. Mannella, Spectroscopic evidence for strong quantum spin fluctuations with itinerant character in YFe$_2$Ge$_2$, Phys. Rev. B 91 (2015) 121102.

\bibitem{che16a}
Jiasheng Chen, Konstantin Semeniuk, Zhuo Feng, Pascal Reiss, Philip Brown, Yang Zou, Peter W. Logg, Giulio I. Lampronti, and F. Malte Grosche, Unconventional Superconductivity in the Layered Iron Germanide YFe$_2$Ge$_2$, Phys. Rev. Lett. 116 (2016) 127001.

\bibitem{zav87a}
I. Yu. Zavalii, V. K. Pecharskii, and O. I. Bodak, Crystal structures of the compounds CuFe$_2$Ge$_2$ and Cu$_{1 \mp x}$Co$_{2 \mp x}$Ge$_2$, Kristallogragiya 33 (1987) 66-69 [Sov. Phys. Crystallogr. 32 (1987) 35-37].

\bibitem{sha15a}
K. V. Shanavas and D. J.  Singh, Itinerant Magnetism in Metallic CuFe$_2$Ge$_2$. PLoS ONE 10 (2015) e0121186. 

\bibitem{may16a}
Andrew F. May, Stuart Calder, David S. Parker, Brian C. Sales, and Michael A. McGuire, Competing magnetic ground states and their coupling to the crystal lattice in CuFe$_2$Ge$_2$, Sci. Rep. 6 (2016) 35325.

\bibitem{lar00a}
A. C. Larson and R. B. Von Dreele, General Structure Analysis System (GSAS), Los Alamos National Laboratory Report LAUR 86-748 (2000). 

\bibitem{kle16a}
Z. Klencz\'ar, MossWinn 4.0 Manual (2016).

\bibitem{pre12a}
C. Prescher, C. McCammon and L. Dubrovinsky, MossA: a program for analyzing energy-domain M\"ossbauer spectra from conventional and synchrotron sources, J. Appl. Cryst. 45
(2012) 329-331.

\bibitem{all16a}
 J. M. Allred,  K. M. Taddei, D. E. Bugaris, M. J. Krogstad, S. H.Lapidus, D. Y. Chung, H. Claus, M. G. Kanatzidis, D. E. Brown, J. Kang, R. M. Fernandes, I. Eremin, S. Rosenkranz, O. Chmaissem, and R. Osborn, Double-Q spin-density wave in iron arsenide superconductors, Nat. Phys. 12  (2016) 493-498. 

\bibitem{bar90a}
J.J. Bara, H.U. Hrynkiewicz, A. Mi{\l}o\'s, A. Szytu{\l}a, Investigation of the crystal properties of RFe$_2$Si$_2$ and RFe$_2$Ge$_ 2$ by X-ray diffraction and M\"ossbauer spectroscopy, J. Less Common Met. 161 (1990) 185-192.

\bibitem{mcg14a}
Michael A. McGuire, Magnetism and Structure in Layered Iron Superconductor Systems, in: Handbook of Magnetic Materials, Volume 22, edited by K. H. J. Buschow, Elsevier, Amsterdam  (2014) 381-463.

\bibitem{blu03a}
S. J. Blundell, C. A. Steer, F. L. Pratt, I. M. Marshall, W. Hayes, and R. C. C. Ward, Detection of magnetic order in the $S = 1$ chain compound LiVGe$_2$O$_6$ using implanted spin-polarized muons, Phys. Rev. B 67 (2003) 224411.

\bibitem{dub09a}
S.M. Dubiel, Relationship between the magnetic hyperfine field and the magnetic moment, J. Alloys Compd. 488 (2009) 18–22. 

\bibitem{iyo16a}
 A. Iyo, K. Kawashima, T. Kinjo, T. Nishio, S. Ishida, H. Fujihisa, Y. Gotoh, K. Kihou, H. Eisaki, and Y. Yoshida, New-Structure-Type Fe-Based Superconductors: CaAFe$_4$As$_4$ (A = K, Rb, Cs) and SrAFe$_4$As$_4$ (A = Rb, Cs), J. Am. Chem. Soc. 138,  (2016) 3410-3415.

\bibitem{mei17a}
W. R. Meier, Q.-P. Ding, A. Kreyssig, S. L. Bud'ko, A. Sapkota, K. Kothapalli, V. Borisov, R. Valentí, C. D. Batista, P. P. Orth, R. M. Fernandes, A. I. Goldman, Y. Furukawa, A. E. B\"ohmer, and  P. C. Canfield, Hedgehog spin vortex crystal in a hole-doped iron based superconductor,  arXiv:1706.01067 (2017).



\end{thebibliography}
\end{document}